\documentclass{sig-alternate-10pt}
\usepackage[margin=1in]{geometry}
\usepackage[numbers]{natbib}
\usepackage{comment}
\usepackage{paralist}
\usepackage{adjustbox}
\usepackage{multirow}
\usepackage{tablefootnote}
\usepackage{subcaption}
\usepackage[normalem]{ulem}
\usepackage{listings}
\usepackage{xcolor}
\usepackage{url}
\usepackage[labelfont=bf, textfont=bf]{caption}
\definecolor{codegreen}{RGB}{0,120,0}
\usepackage{titlesec}
\titleformat{\section}
  {\normalfont\large\bfseries\uppercase}
  {\thesection.}                         
  {0.5em}                                 
  {}
\titleformat{\subsection}
  {\normalfont\large\bfseries}
  {\thesubsection.}
  {0.5em}
  {}
\titlespacing*{\section}{0pt}{6pt plus 2pt minus 1pt}{0pt}
\titlespacing*{\subsection}{0pt}{6pt plus 2pt minus 1pt}{0pt}
\begin{document}

\title{DaiSy: A Library for Scalable Data Series Similarity Search}

\numberofauthors{5} 

\author{
\alignauthor Francesca Del Gaudio\thanks{These authors contributed equally to this work.}\\
       \affaddr{Universit\'e Paris Cit\'e, LIPADE}\\
       \affaddr{F-75006, Paris, France}\\
       \email{\large francescadelgaudio56\\
       @gmail.com}
\alignauthor Manos Chatzakis\footnotemark[1]\\
       \affaddr{Universit\'e Paris Cit\'e, LIPADE}\\
       \affaddr{F-75006, Paris, France}\\
        \email{\large manos.chatzaki\\
        @gmail.com}
\alignauthor Gayathiri Ravendirane\\
       \affaddr{Université de Bordeaux}\\
       \affaddr{Bordeaux, France}\\
       \email{\large gayathiri.ravendirane\\
       @etu.u-bordeaux.fr}
\and
\alignauthor Botao Peng\\
       \affaddr{Chinese Academy of Sciences}\\
       \affaddr{Beijing, China}\\
       \email{\large pengbotao@ict.ac.cn}
\alignauthor Themis Palpanas\\
        \affaddr{Universit\'e Paris Cit\'e, LIPADE}\\
       \affaddr{F-75006, Paris, France}\\
       \email{\large themis@mi.parisdescartes.fr}
}

\maketitle

\begin{abstract}
Exact similarity search over large collections of data series is a fundamental operation in modern applications, yet existing solutions are often fragmented, specialized, or tailored to specific execution environments.
In this paper, we present DaiSy, a unified library for exact data series similarity search that integrates multiple state-of-the-art 
(iSAX-based)
algorithms within a single, coherent framework.
DaiSy is the first library to support exact similarity search across diverse execution environments, including implementations for disk-based, in-memory, GPU-accelerated, and distributed scalable similarity search.
Although designed for data series, DaiSy is also directly applicable to exact similarity search over vector data, enabling its use in a broader range of applications.
The library supports interfaces in both C++ and Python, enabling users to easily integrate its functionality into a variety of tasks.
DaiSy is open-sourced and available at: {\texttt{\url{https://github.com/MChatzakis/DaiSy}}}.
\end{abstract}

\section{INTRODUCTION}
Data series constitute one of the most popular data types, appearing across a wide range of application domains, including finance, astrophysics, neuroscience, engineering, seismology, and many others~\cite{zoumpatianos2018fulfillingtheneed,palpanas2017datamining}.
Such domains typically generate large collections of data series that must be analyzed in order to extract meaningful knowledge~\cite{kashino1999time, ye2009time, boniol2020series2graph, wang2023iedeal, huijse2014computational, raza2015practical}.
Within these data analytics tasks, similarity search serves as a fundamental operation that enables efficient and effective analysis.
Moreover, many similarity search applications are highly sensitive to result accuracy, highlighting the need for similarity search methods that avoid approximation errors, i.e., exact similarity search~\cite{echihabi2018lernaeanhydra}.
In exact query answering, it is crucial to guarantee that the returned data series is indeed the most similar to the query among all data series in the collection.

\noindent \textbf{Scalable Similarity Search.}
The increasing interest in scalable similarity search across different application domains has led to massive research on search time optimization~\cite{echihabi2021newtrendsimsearch, echihabi2022scalablesimsearchtutorial}.
To enable scalable similarity search, numerous indexing methods have been proposed~\cite{palpanas2020dsindexevolution}.
These methods construct index structures over data series collections and employ specialized search algorithms to answer queries.
A large portion of prior work focuses on optimizing the core similarity search algorithms themselves, aiming to reduce query latency and improve throughput~\cite{echihabi2023pros, gogolou2019progressive,kondylakis2019coconutpalm, linardi2018ulisse,fatourou2023fresh}.
At the same time, several approaches extend similarity search to different execution environments based on the available hardware, such as disk-based systems~\cite{peng2018paris}, in-memory CPUs~\cite{peng2020messi}, GPU-accelerated systems~\cite{peng2021sing}, and distributed infrastructures~\cite{chatzakis2023odyssey, yagoubi2017dpisax}.
However, despite this rich body of work, existing methods are largely dispersed, with implementations scattered across different codebases and system designs, which makes practical adoption challenging.

\noindent \textbf{DaiSy.}
We present DaiSy, the first scalable solution for exact data series similarity search that integrates multiple algorithms into a single, unified library.
DaiSy enables efficient similarity search across a wide range of environments and system configurations by supporting scalable algorithms for disk-based, in-memory, GPU-accelerated, and distributed execution.
This is achieved by incorporating representative state-of-the-art 
(iSAX-based)
algorithms for each execution environment: ParIS+~\cite{peng2018paris} for disk-based processing, MESSI~\cite{peng2020messi} for in-memory processing, SING~\cite{peng2021sing} for GPU-accelerated processing, and Odyssey~\cite{chatzakis2023odyssey} for distributed processing.
ParIS+ efficiently supports disk-based similarity search when data do not fit in main memory, MESSI provides highly efficient parallel in-memory search, SING exploits GPU processing to accelerate key search operations 
targetted to CUDA-enabled GPUs
, and Odyssey enables exact similarity search over massive datasets using distributed memory and optimized communication.
Currently, GPU acceleration is available only on CUDA-enabled GPUs. 

All four above algorithms, Paris+, MESSI, SING, and Odyssey, are guaranteed to always return the exact, correct answers. 
The selection of these systems was motivated by the fact that they represent the state-of-the-art approaches for exact similarity search~\cite{echihabi2020hydra2, DBLP:journals/debu/00070P023}.

Finally, we note that although these algorithms were designed with data series in mind, they are equally applicable to general high-dimensional vectors (e.g., deep embeddings), where they exhibit state-of-the-art performance, as well~\cite{echihabi2018lernaeanhydra, echihabi2020hydra2, DBLP:journals/debu/00070P023}. 
This capability is particularly important today, since it constitutes a core component of modern Retrieval-Augmented Generation (RAG) systems for LLMs~\cite{DBLP:conf/wims/EchihabiZP20, DBLP:journals/debu/KrishnaswamyMS24, DBLP:journals/debu/BruchNRV24, sanca2023context, lewis2020rag}.
In addition, approximate vector search applications behind RAG systems usually require some form of training~\cite{chatzakis2025darth,chatzakis2026darthplus} and/or tuning~\cite{lu2026fvsbenchmark}, which requires exact search to obtain groundtruth results.
In such scenarios, DaiSy can significantly accelerate the groundtruth generation process, and represents the current state-of-the-art solution for this task.

DaiSy exposes a unified interface for all supported algorithms and is available in both C++ and Python to facilitate adoption by users.
DaiSy is open sourced and available at: \url{https://github.com/MChatzakis/DaiSy}.

\noindent \textbf{Contributions.}
We summarize our main contributions as follows:

\noindent $\bullet$
We present DaiSy, the first scalable library for exact data series, as well as vector, similarity search that integrates multiple state-of-the-art algorithms within a unified framework.

\noindent $\bullet$
We describe how DaiSy enables efficient data series similarity search across a wide range of configurations and systems, by supporting disk-based, in-memory, GPU-accelerated, and distributed execution environments.

\noindent $\bullet$
We design an extensible similarity search library architecture for DaiSy, accompanied by comprehensive benchmarking frameworks, and we provide both C++ and Python interfaces, facilitating easy adoption across diverse domains and applications.

\noindent $\bullet$
We release DaiSy as an open-source library and make it publicly available to the community through our GitHub repository.

\section{BACKGROUND}

\noindent\textbf{Data Series.} A \emph{data series} $S = \{p_1, \ldots, p_n\}$ is a sequence of $n$ points, where each $p_i = (u_i, t_i)$ represents a value $u_i$ at position $t_i$. 

\noindent\textbf{iSAX Summary.} The \emph{iSAX summary}~\cite{DBLP:conf/kdd/ShiehK08} discretizes a data series by dividing the x-axis into equal segments (represented by their mean) and the y-axis into regions based on normal distribution breakpoints. Each region is assigned a symbolic representation with a specific \emph{cardinality}, enabling hierarchical \emph{iSAX-based index trees}~\cite{palpanas2020dsindexevolution}.

\noindent\textbf{Similarity Search.} Given a collection $\mathcal{C}$ and a query $S$, \emph{similarity search} identifies the top-$k$ series in $\mathcal{C}$ most similar to $S$ according to a distance measure.

\noindent\textbf{Distance Measures.} The \emph{L2 (Euclidean) Distance} is $L2S(T,S) = \sqrt{\sum_{i=1}^{n} (t_i - s_i)^2}$. The distance between iSAX summaries provides a \emph{lower-bound} to the real distance. \emph{Dynamic Time Warping (DTW)} allows non-linear alignments by computing the minimum cumulative cost over all valid warping paths~\cite{keogh2005dtw}.

\subsection{Related Work}
\noindent\textbf{Data Series Indexes.}
Various indices, specific to data series, have been proposed in the literature~\cite{echihabi2021newtrendsimsearch, echihabi2022scalablesimsearchtutorial}. 
DSTree~\cite{wang2013dstree} is an index based on the APCA summarization~\cite{keogh2001apca}. 
The DSTree can adaptively perform split operations by increasing the detail of APCA as needed. 
The iSAX index is based on the SAX summarization, and its extension, iSAX~\cite{DBLP:conf/kdd/ShiehK08}. 
Other iSAX-based indices have been proposed in the literature~\cite{camerra2010isax,peng2021sing, peng2020messi, peng2018paris, yagoubi2017dpisax, chatzakis2023odyssey, kondylakis2019coconutpalm, echihabi2023pros, gogolou2019progressive, echihabi2022hercules, linardi2018ulisse, DBLP:conf/srds/FatourouKPP23, DBLP:journals/vldb/WangWWPW24}, summarized in~\cite{palpanas2020dsindexevolution} and evaluated in~\cite{echihabi2018lernaeanhydra}. 

\noindent\textbf{Similarity Search Libraries.}
\textbf{FAISS}~\cite{douze2025faiss} is a library that implements a wide range of algorithms for approximate vector similarity search.
It is conceptually the most closely related system to DaiSy; however, it targets application scenarios in which the exactness constraint can be relaxed, e.g. Retrieval Augemented Generation (RAG)~\cite{lewis2020rag}. 
FAISS provides a limited functionality for exact similarity search through a naive bruteforce search implementation.
In addition, \textbf{UCR Suite}~\cite{rakthanmanon2012ucrsuite} and \textbf{TSSEARCH}~\cite{folgado2022tssearch} are libraries for subsequence similarity search in C++ and Python respectively.
They both tackle a similar variant of similarity search, but they do not provide various different methods to perform indexing or search.
\textbf{AEON}~\cite{middlehurst2024aeon} and \textbf{TSLEARN}~\cite{tavenard2020tslearn} are machine learning oriented Python libraries for data series processing, with strong focus on analytics tasks. 
Although they provide some functionalities for computing distances between data series, they are not libraries for scalable data series similarity search and they do not expose any interface for indexing or search.

\section{SYSTEM DESIGN}

DaiSy's architecture follows a layered design centered on a C++ core that implements indexing, distance computation, and search functionality, while exposing the same conceptual model through a Python interface. Foundational primitives provide distance evaluation, data access, and reusable lower bounds, while an index layer encapsulates the iSAX index as an internal service~\cite{camerra2010isax}. 
On top of these, multiple execution models realize exact similarity search under different assumptions on data placement and available resources. 
The same abstractions are exposed in both C++ and Python, ensuring a uniform user-facing interface independent of the execution back ends enabled at build time.

\subsection{Architecture}
\label{sec:architecture}

\begin{figure*}
    \centering
    \includegraphics[width=0.95\textwidth]
    {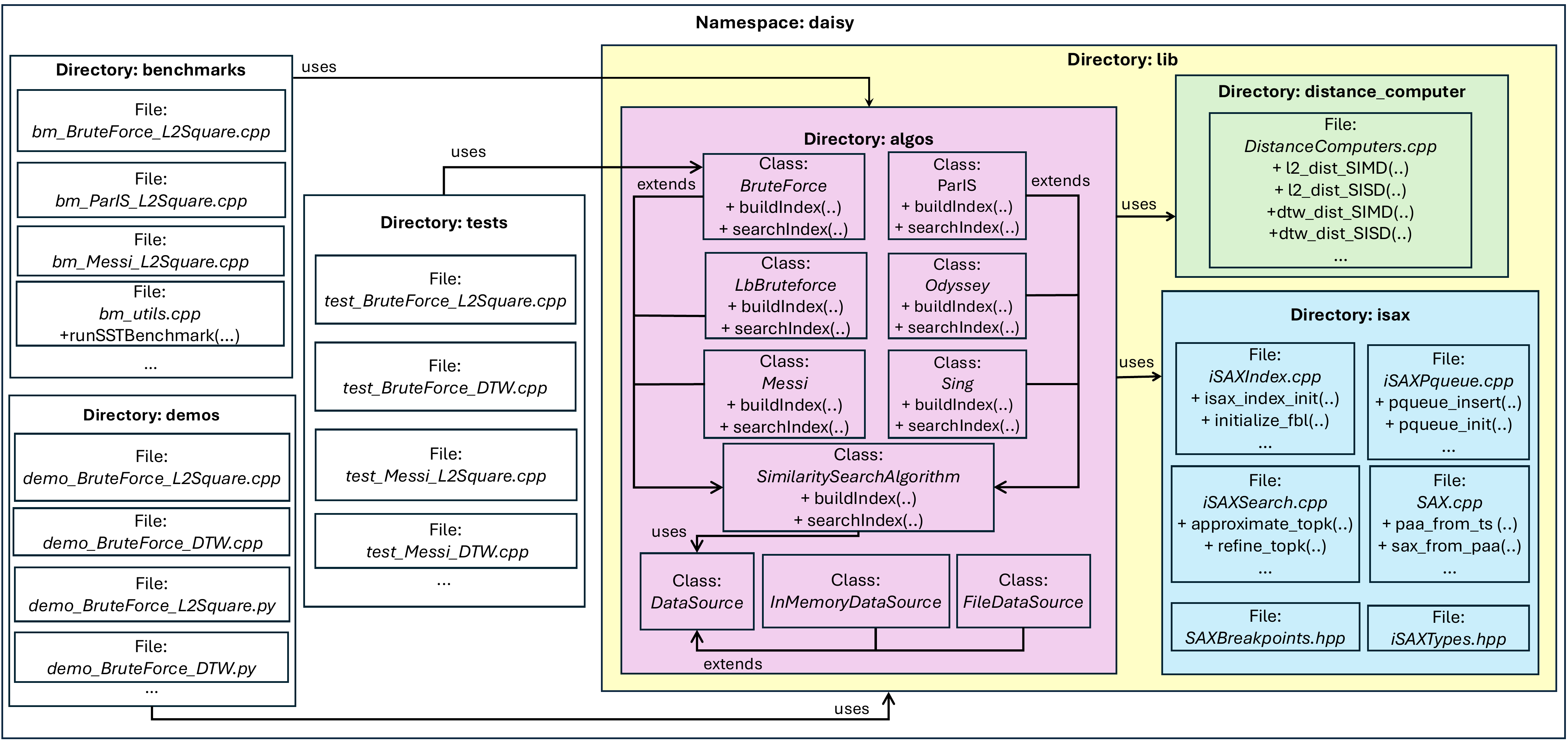}
     \caption{Component Diagram of DaiSy.}
    \label{fig:architecture}
\end{figure*}

\noindent\textbf{Design principles and abstractions.}
DaiSy is structured around a set of design principles that emphasize modularity,  enforcing a strict separation of concerns along three orthogonal dimensions. 
Distance computation is encapsulated in a dedicated component responsible for exact distance evaluation and lower bounds, independent of data storage and execution strategy. 
Data access is handled through an explicit adapter-based abstraction~\cite{DBLP:conf/cit/Al-ObeidallahAK21}, which decouples indexing and search logic from the physical layout and location of the data and enables support for heterogeneous data sources, including in-memory and on-disk representations. 
Execution strategy is treated as a separate concern, while search algorithms are implemented as interchangeable components exposing a common interface.\\
\noindent\textbf{Core primitives layer.}
The core primitives layer provides the shared foundation on which all indexing structures and search algorithms in DaiSy are built. 
It deliberately implements no index or search strategy; instead, it factors out functionality common across algorithms and execution models to avoid duplication.
Distance-related logic is encapsulated in a unified abstraction supporting exact evaluation and lower bounds, and is independent of data placement and execution strategy. 
Data access is exposed through a stable adapter-based interface that models datasets as streams of data series, keeping higher-level components agnostic to in-memory or on-disk storage. 
The layer also provides shared reduced representations and lower bounds enabling safe pruning while preserving correctness, forming a stable, metric- and storage-agnostic substrate for higher layers.\\
\noindent\textbf{Index layer.}
The index layer encapsulates the iSAX index as an internal service that provides structured access to data series data, without prescribing a search strategy. 
Index construction is encapsulated by a single configuration object that defines representation parameters, buffer sizes, and storage mode.
Importantly, the index layer does not implement search logic. 
By treating iSAX as a service rather than as an algorithm, DaiSy decouples index management from execution strategy and allows the same index implementation to be reused across multiple execution models.\\
\noindent\textbf{Similarity Search layer.}
The similarity search layer in DaiSy is organized by execution model rather than by individual algorithm implementations. 
Each execution model defines how exact similarity search is carried out under specific assumptions about data placement and available resources, while reusing the same core primitives and index services. 
Taken together, these models position DaiSy as an execution framework that instantiates a single conceptual search pipeline through multiple back ends.
DaiSy supports a variety of similarity search algorithms in this layer, covering use cases in disk-based, in-memory, GPU-accelerated and distributed environments.
ParIS+~\cite{peng2018paris} is an algorithm designed for disk-based exact similarity search, and its tailored for situations where the dataset size exceeds the main memory capacity of the system.
MESSI~\cite{peng2020messi} is an exact similarity search algorithm that operates entirely in memory and relies on an iSAX index to guide candidate selection and refinement. 
It follows a multi-phase search pipeline that combines index-guided exploration with exact refinement while preserving exactness.
The SING index~\cite{peng2021sing} leverages GPUs to accelerate the search procedure. 
It is organized into two main stages: in-memory index construction and multi-phase query answering with GPU offloading.
Odyssey~\cite{chatzakis2023odyssey} is a distributed framework for exact similarity search built on MPI, designed for multi-node environments where datasets can be accommodated in distributed memory. 
DaiSy also supports two exhaustive search implementations. \\
The first one, \emph{Bruteforce}, simply performs all real distance computations over the raw data. 
The second one, \emph{LbBruteforce}, applies lower-bounding, thus, avoiding real distance computations when the lower bound is larger than the BSF distance.

\subsection{Distance Measures}
DaiSy supports L2 Squared and DTW for exact similarity search, both exposed through a unified distance computer abstraction.
Our library supports both optimized SIMD and scalar implementations, to ensure compatibility with different hardware configurations.
Distance computations support early termination via an optional upper bound, allowing evaluation to stop once the partial distance exceeds the current best-so-far value~\cite{rakthanmanon2012ucrsuite}.
In addition, DaiSy provides exact lower-bound functions for both measures.

\subsection{Z-normalization}
DaiSy supports both z-normalized (i.e., each data series, or vector, has mean $0$ and standard deviation~$1$) and non z-normalized data.
Though, the algorithms are more suited to Z-normalized data (we use the default iSAX breakpoints~\cite{DBLP:conf/kdd/ShiehK08}, which have been optimized for z-normalized data).

\section{USING DAISY}

\subsection{API and Tooling}
DaiSy exposes a simple API to enable easy use in both C++ and Python. 
\begin{compactitem}
    \item \texttt{buildIndex.} 
    Initializes and builds the index structure for the selected algorithm.
    \item \texttt{searchIndex.}
    Performs the top-k index search for the given queries, returning the most similar data series from the index.
\end{compactitem}
Furthermore, reproducibility and systematic evaluation are supported through a set of shared utilities, demos, benchmarks, and tests. 
Demos are provided in both C++ and Python and exercise the same pipelines exposed to end users. 
Benchmarks enable controlled performance evaluation across execution models and configurations, while tests validate correctness and integration. 
All tooling components consume the same APIs as the core library, ensuring consistency between usage, evaluation, and experimentation.

\subsection{Hyperparameter Setting}
As expected, the methods implemented in DaiSy involve a number of algorithm-specific hyperparameters.
The library allows experienced users to explicitly configure all relevant hyperparameters for each supported algorithm.
At the same time, DaiSy does not require the user to manually specify these parameters, as it initializes them to the values recommended in the original publications of the corresponding algorithms.

\subsection{Algorithm Selection}

\begin{figure}[tb]
    \centering
    \includegraphics[width=0.99\linewidth]
    {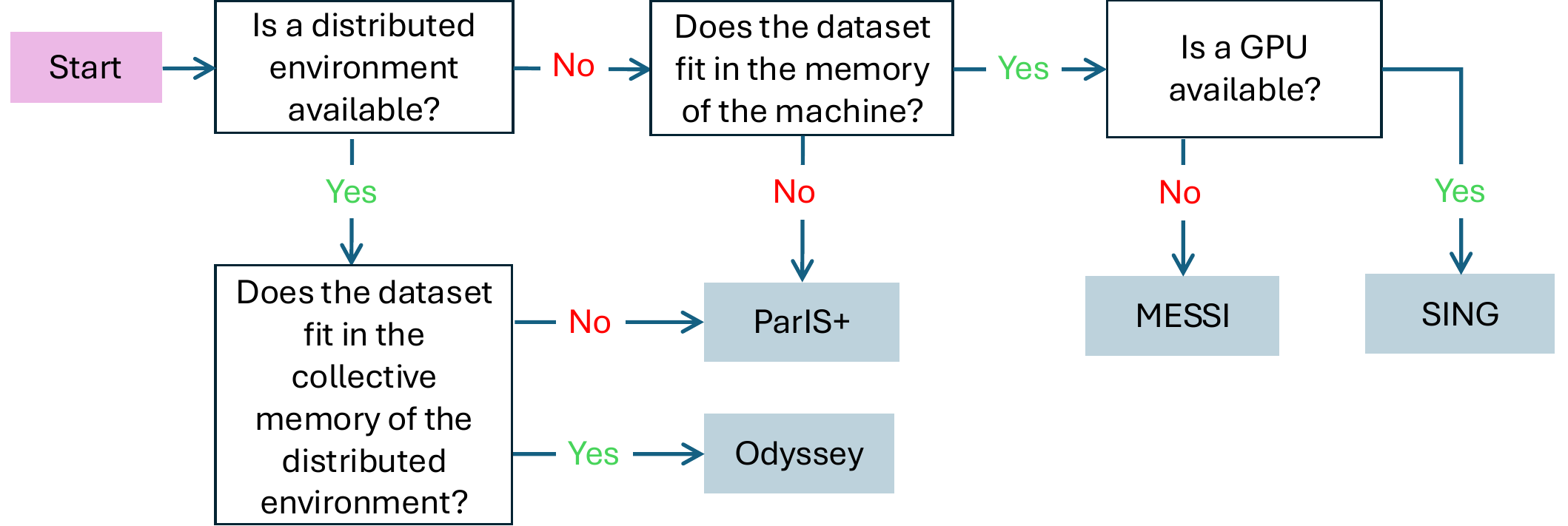}
    \caption{Algorithm selection decision tree of DaiSy.}
    \label{fig:daisy_decision_tree}
    \vspace{-5mm}
\end{figure}

As we demonstrated, DaiSy provides efficient similarity search algorithms tailored for different execution environments.
Figure~\ref{fig:daisy_decision_tree} summarizes the algorithm selection logic based on dataset size and available resources. 
When the dataset fits in main memory, DaiSy first considers whether a distributed environment is available and, if so, selects Odyssey to exploit distributed memory. 
In the absence of a distributed setting, the choice is driven by the available hardware accelerators: SING is used when a GPU is present, while MESSI is selected for CPU-based in-memory execution. 
When the dataset does not fit in main memory, DaiSy resorts to ParIS+ for disk-based execution.

\begin{figure}
    \hspace*{-0.5cm}
    \centering
    \begin{adjustbox}{max width=0.30\textwidth}
        \includegraphics{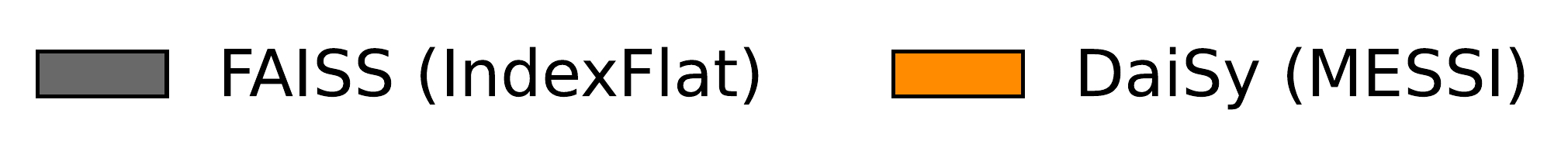}
    \end{adjustbox}

    \begin{minipage}[t]{0.49\textwidth}
        \centering

        \begin{subfigure}[t]{0.49\textwidth}
            \centering
            \includegraphics[width=\textwidth]{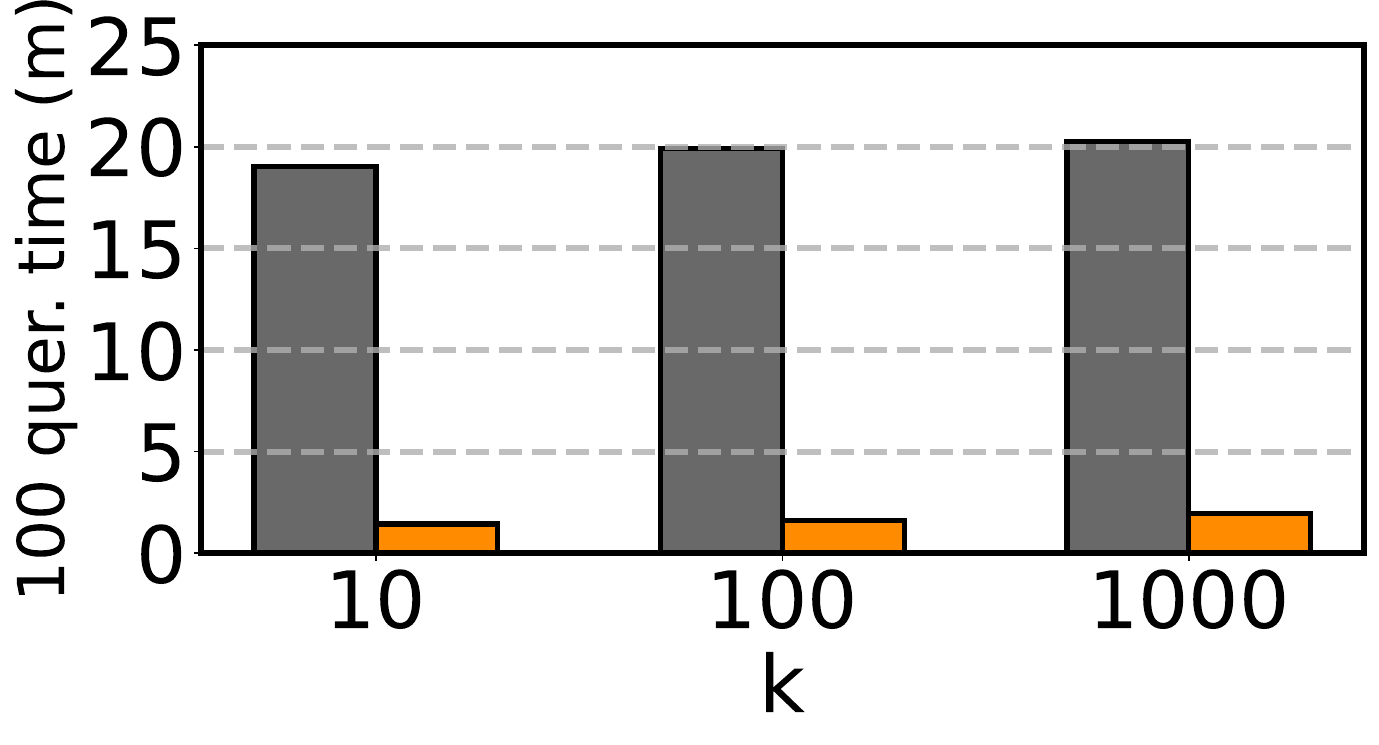}
        \caption{Deep100M
        }
        \end{subfigure}
        \begin{subfigure}[t]{0.49\textwidth}
            \centering
            \includegraphics[width=\textwidth]{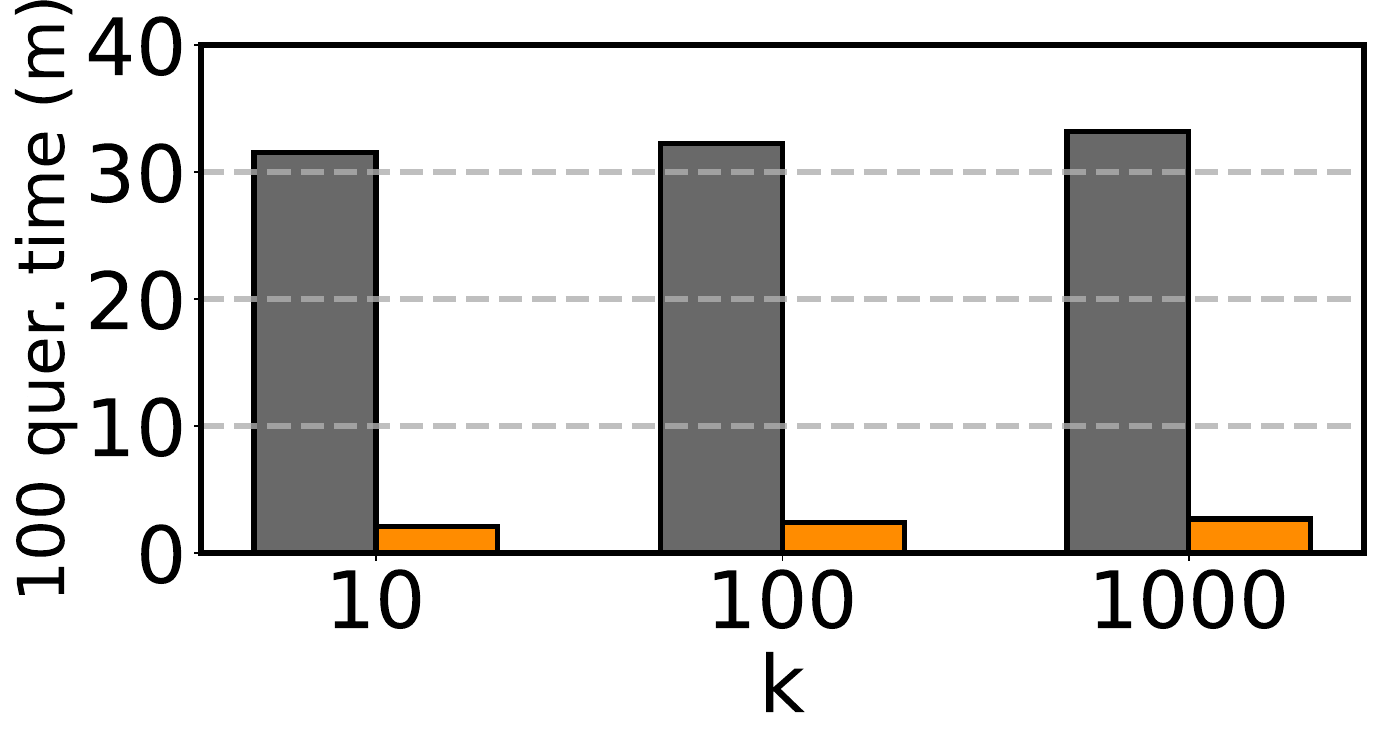}
            \caption{Seismic100M}
        \end{subfigure}
    \caption{Query answering time for 100 queries when varying k (48 hyperthreads).} 
    \label{fig:varying-k}
    \end{minipage}
\end{figure}

We would like to stress that, apart from data series, DaiSy is equally applicable to vector data, as well, where it exhibits state-of-the-art performance. 
For example, DaiSy is the solution of choice for applications that need exact answers in vector search, or for analysis tasks that involve vector approximate search and need to compute the groundtruth answers. 
As evidence, we present the results of the following experiment (rigorous experimental evaluations can be found in previous studies~\cite{echihabi2018lernaeanhydra, echihabi2020hydra2, chatzakis2023odyssey}). 
We conduct an in-memory evaluation against FAISS-IndexFlat, the most widely used library for such tasks, focusing on exact similarity search for two datasets: Deep100M~\cite{babenko2016deep} (100 million 96-dimensional image embeddings) and Seismic100M~\cite{seismic2018} (100 million 256-dimensional seismic data series).
IndexFlat is the FAISS solution used to compute exact answers: it stores the full vectors and performs an exhaustive search.
Both libraries are compiled with identical settings using \texttt{g++ 11.4.0}, configured for in-memory execution, and execution times are measured using the GoogleBenchmark~\footnote{\url{https://github.com/google/benchmark}} framework.
Figure~\ref{fig:varying-k} reports the total query execution time for 100 queries on each dataset, when varying $k$ between $10$-$1000$, using 48 hyperthreads.

The results show that DaiSy-MESSI consistently outperforms FAISS-IndexFlat, calculating the exact answers for the entire workload $12\times$ faster for the Deep100M vector collection. 
The same observation is true for the Seismic100M data series collection, where DaiSy-MESSI is $14\times$ faster. Speed-up increases with the number of threads.
Similar results hold for other vector datasets, as well.

\subsection{Examples}
We provide two examples of using DaiSy in C++ and Python to demonstrate the convenience our library provides for similarity search.
Without loss of generality, in our examples we use the MESSI algorithm and the L2 Squared distance.
\begin{lstlisting}[language=C++,basicstyle=\ttfamily\scriptsize,numbers=none]
#include <vector>
#include "daisy/daisy.hpp"
int main() {
    daisy::idx_t d    = 256;    // dimensionality
    daisy::idx_t n_db = 10000;  // dataset size
    daisy::idx_t n_q  = 100;    // queries size

    // Load or generate input data series.
    float* db = /* load or generate database series */;
    float* q  = /* load or generate query series */;

    // Build index using squared Euclidean distance.
    daisy::Messi messi(daisy::DistanceType::L2_SQUARED);
    messi.buildIndex(db, n_db, d);

    // Execute exact k-NN similarity search.
    const daisy::idx_t k = 5;
    std::vector<daisy::idx_t> I(static_cast<size_t>(n_q)*k);
    std::vector<float>        D(static_cast<size_t>(n_q)*k);
    messi.searchIndex(q, n_q, k, I.data(), D.data());
    
    return 0; }
\end{lstlisting}
\noindent
\begin{lstlisting}[
    language=Python,
    basicstyle=\ttfamily\scriptsize,
    numbers=none
]
from daisy import DistanceType, Messi

d = 256       # dimensionality
n_db = 100000 # dataset size 
n_q = 1000    # queries size

# Load or generate input data.
db = ... # load or generate database series 
q  = ... # load or generate query series 

# Build the index using squared Euclidean distance.
index = Messi(DistanceType.L2_SQUARED)
index.buildIndex(db)

# Execute exact k-NN similarity search.
k = 5
I, D = index.searchIndex(q, k)
\end{lstlisting}

\section{FUTURE WORK}
DaiSy serves as the foundation for several planned future efforts aimed at further improving and extending the library.
In particular, we plan to incorporate additional algorithms for exact similarity search, such as DumpyOS~\cite{DBLP:journals/vldb/WangWWPW24}, 
including methods that do not rely on the iSAX index, such as Hercules~\cite{echihabi2022hercules} and SOFA~\cite{DBLP:conf/icde/SOFA}, which is enabled by the modular architecture of DaiSy.
We also plan to extend the library to support range queries~\cite{echihabi2018lernaeanhydra}, as well as subsequence similarity search~\cite{rakthanmanon2012ucrsuite, DBLP:journals/access/FengWWW20, linardi2018ulisse, DBLP:journals/pvldb/XiongZWHWW24}, streaming data series~\cite{DBLP:journals/vldb/KondylakisDZP19}, and early termination~\cite{gogolou2019progressive, echihabi2023pros, chatzakis2025darth}.
We also intend to introduce automatic hyperparameter tuning capabilities for the supported algorithms, such as Bayesian optimization~\cite{daulton2020mobo}.
Our immediate priorities will focus on integrating non-iSAX-based systems and on supporting range-based queries.

\section{CONCLUSIONS}
We present DaiSy, a novel library for scalable exact similarity search on large collections of data series, as well as general high-dimensional vectors (e.g., deep embeddings).
It supports efficient search algorithms across a wide range of hardware configurations. 
We open source DaiSy and provide carefully designed C++ and Python interfaces.

\section*{Acknowledgments}
Work supported by EU Horizon projects ARMADA ($101168951$), TwinODIS ($101160009$), DataGEMS ($101188416$), and $Y \Pi AI \Theta A$ \& NextGenerationEU project HARSH ($Y\Pi 3TA-0560901$) that is carried out within the framework of the National Recovery and Resilience Plan “Greece 2.0” with funding from the European Union – NextGenerationEU.
Manos Chatzakis is supported with a PhD Scholarship from the Onassis Foundation.

\begin{small}
\section{REFERENCES}
\renewcommand{\refname}{\setlength{\itemsep}{-2,5pt}}
\vspace{-15pt}
\begingroup
    \scriptsize 
    \linespread{0.83}\selectfont
    \bibliographystyle{abbrv}
    \bibliography{references}
\endgroup
\end{small}

\end{document}